\documentclass[letterpaper]{article} 
\usepackage{aaai24}  
\usepackage{times}  
\usepackage{helvet}  
\usepackage{courier}  
\usepackage[hyphens]{url}  
\usepackage{graphicx} 
\usepackage{natbib}  
\usepackage{caption} 
\usepackage{algorithm}
\usepackage{algorithmic}
\usepackage{amsmath,amssymb,amsthm,}
\newtheorem{theorem}{Theorem}
\newtheorem{definition}{Definition}
\newtheorem{proposition}{Proposition}
\newtheorem{lemma}{Lemma}
\newtheorem{example}{Example}

\newcommand*{\boldone}{\text{\usefont{U}{bbold}{m}{n}1}}
\usepackage{verbatim}
\usepackage{newfloat}
\usepackage{listings}
\DeclareCaptionStyle{ruled}{labelfont=normalfont,labelsep=colon,strut=off} 
\floatstyle{ruled}
\newfloat{listing}{tb}{lst}{}
\floatname{listing}{Listing}
\title{The Complexity of Computing Robust Mediated Equilibria in Ordinal Games}
\author{
Vincent Conitzer
}
\affiliations{
    Carnegie Mellon University\\

    Pittsburgh, PA, USA\\
    conitzer@cs.cmu.edu
}

\usepackage{bibentry}

\begin{document}

\maketitle

\begin{abstract}
Usually, to apply game-theoretic methods, we must specify utilities
precisely, and we run the risk that the solutions we compute are not
robust to errors in this specification.  Ordinal games provide an
attractive alternative: they require specifying only which outcomes
are preferred to which other ones.  Unfortunately, they provide little
guidance for how to play unless there are pure Nash equilibria;
evaluating mixed strategies appears to fundamentally require cardinal
utilities.

In this paper, we observe that we can in fact make good use of mixed
strategies in ordinal games if we consider settings that allow for
folk theorems.  These allow us to find equilibria that are robust, in
the sense that they remain equilibria no matter which cardinal
utilities are the correct ones -- as long as they are consistent with
the specified ordinal preferences.  We analyze this concept and study
the computational complexity of finding such equilibria in a range of
settings.
\end{abstract}

\section{Introduction}

Game theory (see,
e.g.,~\cite{Fudenberg91:Game_theory,Shoham09:Multiagent}) is the study
of rational behavior in the presence of other agents that also behave
rationally, but with possibly different utility functions.  As such,
it should be useful for the design and analysis of systems of multiple
self-interested agents in AI, and certainly for many purposes it is.
But one limitation is that game theory generally requires the
specification of precise utilities.  In games with clear rules -- say,
poker -- this is not a problem. But out in the real world, we often do
not know the players' precise utilities.  One method for dealing with
uncertainty about utilities in game theory is to use {\em Bayesian
  games}, in which there is a common prior distribution over the
agents' utilities.  However, determining an accurate prior
distribution is no less demanding a task.  More practical are methods
that assume relatively little about the utility functions and are
inherently {\em robust} (e.g.,~\cite{Pita10:Robust,Gan23:Robust}.
Such robustness can also bring significant safety benefits to AI
systems, especially as they interact with a complex human world.
Otherwise, if agents fail to correctly estimate each other's utility
functions, the result can be disastrous.

For example, consider the game on the left in Figure~\ref{fi:intro}.
\begin{figure}
\begin{center}
\begin{tabular}{| c | c | c |}\hline
8, 9 & 9, 9 & 0, 0\\  \hline
7.8, 6 & 9.1, 6 & 6, 7\\  \hline
8.1, 6 & 8.8, 6 & 7, 7\\  \hline
\end{tabular} \ \ \ \ \
\begin{tabular}{| c | c | c |}\hline
8, 9 & 9, 9 & 0, 0\\  \hline
7.9, 6 & 9.2, 6 & 6, 7\\  \hline
8.2, 6 & 8.9, 6 & 7, 7\\  \hline
\end{tabular} \ \ \ \ \
\end{center}
\caption{Left: game with an equilibrium with utilities $8.5,9$.
  Right: only the equilibrium with utilities $7,7$ survives.}
\label{fi:intro}
\end{figure}
One Nash equilibrium is for player 1 to play Top and player 2 to mix
50-50 between Left and Center, for utilities of $(8.5,9)$; another is
(Bottom, Right) for utilities of $(7,7)$. It seems natural to play the
former equilibrium given the higher utilities.  However, in the
slightly modified game on the right of Figure~\ref{fi:intro}, only the
latter equilibrium survives.  (This follows by iterated elimination of
strictly dominated strategies, because mixing 50-50 between Middle and
Bottom now strictly dominates Top.)  Now, if player 1 thinks the game
is the left one and player 2 thinks it is the right one, they may well
play the disastrous outcome (Top, Right).\footnote{In particular, note
  that player 2 cannot distinguish the games based on knowing his own
  utilities.}

{\em Ordinal
  games}~\cite{Cruz00:Ordinal,Bonanno08:Syntactic,Gafarov15:Ordinal}
constitute one appealing framework for robustness.  In such a game, it
is assumed we know nothing more than ordinal preferences over the
outcomes, e.g., we know that player 1 weakly prefers outcome 1 to
outcome 2 -- but not by how much.  Indeed, across the two games in
Figure~\ref{fi:intro}, each player's ordinal comparisons of outcomes
are the same, so they correspond to the same ordinal game; and
abstracting to this ordinal game would focus attention on the (Bottom,
Right) pure equilibrium.

To illustrate: it is generally safe to assume that a student will
prefer a grade of A (``excellent'') to a grade of B (``good''), and a
grade of B to a grade of C (``average''); but the (cardinal) utility
the student gets for each of these outcomes is generally unknown.  A
student that requires at least a B to remain in good standing in the
program may have utility $u(\text{A})=1, u(\text{B})=.99,
u(\text{C})=0$; a student that will be applying to another highly
competitive program that requires all A grades may have utility
$u(\text{A})=1, u(\text{B})=.01, u(\text{C})=0$; and a student that is
in both situations at the same time may have utility $u(\text{A})=1,
u(\text{B})=.5, u(\text{C})=0$.  We often do not know which of these
situations the student is in, and in certain circumstances the
difference would affect the student's behavior (will the student do
the ``safe'' course project that is sure to lead to a B, or the
``risky'' project that might result in either an A or a C?) -- but we
can yet be confident that A $\succeq$ B $\succeq$ C.

As illustrated by this example, a significant downside of this
framework, and one that has likely limited the study and use of it, is
that it generally tells us little about how a player assesses {\em
  distributions} over outcomes.  Thus, we can identify pure Nash
equilibria of such games, but there is little to say about mixed Nash
equilibria, because in mixed Nash equilibria, generally probabilities
are chosen carefully to leave the other player exactly indifferent
between several options -- which cannot be done without access to
cardinal utilities.

The insight of the current paper is that this is no longer so when we
consider richer solution concepts, ones that allow for {\em folk
  theorems}.  Folk theorems can sustain a variety of behavior,
including cooperative behavior, in equilibrium, by the threat of
punishing players who deviate.  As such, they play a key role in the
nascent field of {\em cooperative
  AI}~\cite{Dafoe21:Cooperative,Conitzer23:Foundations}.  Robustness
is especially important in the context of folk theorems, because
failing to estimate accurately whether a particular punishment
strategy is sufficient to sustain a particular cooperative behavior
can result in the cooperation falling apart, and, worse, in triggering
very damaging punishment strategies that were never meant to actually
be played (i.e., they were supposed to be {\em off the path of play}
in equilibrium, meaning that in the equilibrium we had in mind, the
punishment states of the game would never actually be reached).

{\bf Background on folk theorems.} The traditional setting for the
folk theorem is infinitely repeated games, in which players care
either about their limit average payoff, or have a discount factor
that tends to $1$.  To establish the folk theorem, we use Nash
equilibria (of the repeated game) consisting of two components: (1) a
distribution $p$ over action profiles in the game that is played in
equilibrium (``on the path of play'' or simply ``on-path''), in the
sense that the players rotate through these action profiles in a
pre-specified way that results in each of those action profiles being
played a fraction of the time that corresponds to the probability of
that action profile under $p$; and (2) for each player $i$, a
distribution $q_{-i}$ over action profiles for the other players $-i$,
to ``punish'' player $i$ for deviating (if $i$ ever deviates),
designed to minimize $i$'s utility when best-responding to this
profile.  For example, in the games of Figure~\ref{fi:intro}, we could
set $p$ to always play (Top, Center), $q_{-1}$ to always play Right,
and $q_{-2}$ to always play Bottom; this is an equilibrium because
each player prefers (Top, Center) to the best response to punishment,
which is (Bottom, Right) in both cases -- and the ordinal information
of these games is sufficient to establish that.

When there are 3 or more players, there is a question of whether the
players other than $i$ are able to correlate their actions (say,
through private communication among themselves) to punish $i$; we
assume here that they can.  Indeed, it is already well known that
equilibria of repeated games (with cardinal utilities) are, via the
folk theorem, easier to compute than equilibria of single-shot games,
in 2-player games but also in 3$^+$-player games if such correlated
punishment is
allowed~\cite{Littman05:Polynomial,Kontogiannis08:Equilibrium} --
whereas it is computationally hard to do so with 3 players if
correlated punishment is not allowed~\cite{Borgs10:Myth}, though in
practice such equilibria can still often be found
fast~\cite{Andersen13:Fast}.

Another setting that allows for a folk theorem is that of {\em
  mediated equilibrium}~\cite{Monderer09:Strong} (see
also~\cite{Kalai78:Arbitration,Rozenfeld07:Routing,Kalai07:Commitment}).
In mediated equilibrium, we introduce an additional, disinterested
player called the mediator.  Each player (other than the mediator)
first chooses (simultaneously) whether to cede control over its choice
of action to the mediator, or to keep control over its action.  Then,
the mediator and all players who did not cede control play
simultaneously.  The mediator plays according to a fixed strategy,
which takes into account which players ceded control to it, and
therefore can use the actions of players who ceded control to punish
the players that did not. Since here we consider only unilateral
deviations, it suffices to consider only the case of punishing a
single player $i$.  Note that in the setting of mediated equilibrium,
it is entirely natural that the punishment strategy allows for the
correlation of the actions of the players other than $i$, because the
mediator at this point controls all these actions and can naturally
correlate them with one joint distribution $q_{-i}$.

Another relevant concept is that of {\em program
  equilibrium}~\cite{Tennenholtz04:Program} (see
also~\cite{Rubinstein98:Modeling}), in which the players write
programs that will play on their behalf, and these programs can read
each other, so that they can detect deviation by the other player
ahead of play and instantly punish it.  Program equilibrium appears
closely related to mediated equilibrium, and the two concepts have
been conjectured to be equivalent in which outcomes can be implemented
by them~\cite{Monderer09:Strong}.  We will not resolve the exact
relationship between these concepts in this paper, but this conjecture
suggests our results may also apply to the concept of program
equilibrium.

{\bf Our results.} We wish to construct folk-theorem-style equilibria
defined by $p$ and $(q_{-i})_{i}$, for ordinal games.  We want it to
be the case that for {\em every} cardinal utility function consistent
with the ordinal constraints, each player $i$ weakly prefers play
under $p$ to play under $q_{-i}$ (when best-responding to $q_{-i}$).
By the above, such equilibria can be sustained as equilibria of both
repeated games and mediated games; and at least under some
assumptions, equilibria with this structure are without loss of
generality.  We postpone analysis of such assumptions to the end of
the paper.

Technically, it is helpful to consider these questions in the language
of {\em pre-Bayesian games} (see~\cite{Ashlagi06:Resource} and
references therein), which are Bayesian games without any prior
distribution.  In a pre-Bayesian game, at the time of play, every
player has access to its own type, which encodes its utility function
over outcomes.  The set of types is typically restricted.  Ordinal
games can be considered a special case of pre-Bayesian games in which
for each player, the set of possible types is restricted to those that
encode utilities that satisfy the given ordinal constraints.  Here,
the set of each player's types is generally a continuum.
Nevertheless, we will first show how to solve our problem for an
arbitrary pre-Bayesian game with a {\em finite} number of types, as a
linear program, and then extend that to ordinal games.  We will first
extend it to games with total orders for the players, and then to
games with partial orders.  We proceed to show that there are
limitations as well: specifically, for a richer language in which we
are uncertain about which ordering constraints hold but know that a
certain logical formula regarding such ordering constraints must hold
true, the problem becomes computationally hard to solve.  We conclude
by discussing how robust our notion of robust equilibrium is to
modeling assumptions.

\section{Definitions and Notation}

In this paper, we study $n$-player simultaneous-move games.  To
minimize ambiguity, we use ``she'' for player 1, ``he'' for player 2,
and ``it'' for an unspecified player.  Player $i$ has action set
$A_i$; we let $A = A_1 \times A_2 \times \ldots A_n$ and use the
standard notation $A_{-i} = A_1 \times A_2 \times \ldots \times
A_{i-1} \times A_{i+1} \times \ldots \times A_n$ to denote action
profiles of players other than $i$.  There is a set of possible
outcomes $O$ and an outcome selection function $o: A \rightarrow O$.
We use the formalism of pre-Bayesian games, which means that every
player $i$ has a set of possible types $\Theta_i$, and a utility
function $u_i: \Theta_i \times O \rightarrow [0,1]$, normalizing
utilities to lie between $0$ and $1$.  We use $u_i(\theta_i, a)$ as a
shorthand for $u_i(\theta_i, o(a))$.  Given our focus on ordinal
games, in most of the paper, the set $\Theta_i$ is constructed by
ordinal (preference) constraints.  For example, $o \succeq_i o'$
indicates that for every $\theta_i \in \Theta_i$, it is the case that
$u_i(\theta_i, o) \geq u_i(\theta_i, o')$; that is, regardless of the
type, $i$ weakly prefers $o$ to $o'$.  When working with ordinal
constraints, the set of types consists of exactly those types that
satisfy all the constraints.  We are now ready to define our solution
concept.

\begin{definition}
A {\em mediated profile} consists of a probability distribution $p: A
\rightarrow [0,1]$ to play on the path of play in equilibrium, and,
for every player $i$, a probability distribution $q_{-i}: A_{-i}
\rightarrow [0,1]$ for punishing player $i$ for deviating. A mediated
profile is a {\em robust mediated equilibrium} (or simply {\em robust
  equilibrium}) if for every player $i$, every type $\theta_i \in
\Theta_i$, and every action $a_i'$, we have
$$\sum_{a \in A} p(a) u_i(\theta_i, a) \geq \sum_{a_{-i} \in A_{-i}}
q_{-i}(a_{-i}) u_i(\theta_i, a_i', a_{-i})$$ That is, every player is
at least as well off under $p$ as it would be deviating to any
alternative course of action $a_i'$ (triggering punishment from the
mediator / the other players).
\label{def:eq}
\end{definition}

We discuss examples in the next section.  The above definition is
natural in the setting of mediated
equilibrium~\cite{Monderer09:Strong}, as in that setting, a player can
only deviate once to a single action and the mediator will instantly
coordinate others' actions to inflict punishment.  In the context of
infinitely repeated games, it is easy to see that the condition in
Definition~\ref{def:eq} is at least sufficient to prevent deviation,
but it is less obvious that it is necessary; and in fact, whether it
is necessary depends on further assumptions regarding observability of
the deviator's actions, as well as persistence of the deviator's type
over time.  We defer discussion of this to the penultimate section of
the paper.

We will be interested in the following computational problems.  In
each of these, the description of a player's type space may be: a
finite set of types; a total order over outcomes; a partial order over
(distributions over) outcomes; or expressed in the richer language
that can express uncertainty over ordering constraints using logical
connectives.

\noindent {\sc Existence-of-robust-equilibrium (EORE):} Given a
pre-Bayesian game, does it have a robust equilibrium?

\noindent {\sc Supported-in-robust-equilibrium (SIRE):} Given a
pre-Bayesian game and a distinguished action profile $a^*$, does there
exist a robust equilibrium with $p(a^*)>0$?

\noindent {\sc Attainable-as-robust-equilibrium (AARE):} Given a
pre-Bayesian game and a probability distribution $p^*: A \rightarrow
[0,1]$, does there exist a robust equilibrium with $p=p^*$?

\noindent {\sc Objective-maximization-in-robust-equilibrium (OMIRE):}
Given a pre-Bayesian game, an objective function $g: A \rightarrow
[0,1]$, and a target value $T$, does there exist a robust equilibrium
such that $\sum_{a \in A} p(a)g(a) \geq T$?

For each of these, when we give a positive result, we do so by giving
an algorithm that computes $p$ and the $(q_{-i})_i$.

\section{Examples and Basic Results}

We first show that not all ordinal games (even with total orders) have
robust equilibria. Consider the game in Figure~\ref{fi:MP1}. (It is
helpful to illustrate ordinal games with possible cardinal utilities,
as is done on the right of the figure, but the utilities could also be
different; e.g., every number could be independently increased or
decreased by up to $1/2$, and we would still end up with utilities
consistent with the ordinal game.)
\begin{figure}
\begin{center}
\begin{tabular}{| c | c |}\hline
$o_{11}$ & $o_{12}$\\  \hline
$o_{21}$ & $o_{22}$\\  \hline
\end{tabular} \ \ \ \ \
\begin{tabular}{| c | c |}\hline
4, 1 & 2, 3\\  \hline
1, 4 & 3, 2\\  \hline
\end{tabular}
\end{center}
\caption{Left: together with the orders $o_{11} \succeq_1 o_{22}
  \succeq_1 o_{12} \succeq_1 o_{21}$ and $o_{21} \succeq_2 o_{12}
  \succeq_2 o_{22} \succeq_2 o_{11}$, this is an ordinal game.  Right:
  example utilities satisfying these constraints.}
\label{fi:MP1}
\end{figure}
This is a type of matching pennies game, except one where players have
possibly different utilities for their two ``winning'' outcomes, as
well as for their ``losing'' outcomes.

\begin{proposition}
Figure~\ref{fi:MP1}'s game has no robust equilibrium.
\label{prop:nonexistence}
\end{proposition}
\begin{proof}
First, we argue that $o_{11}$ cannot be played with positive
probability in equilibrium (on path).  This is because it is possible
that player $2$ assigns utility $0$ to it and utility $1$ to all the
other three outcomes; and if so, if $o_{11}$ received positive
probability, player $2$ would be better off deviating and always
playing Right, guaranteeing himself a utility of $1$ regardless of the
punishment strategy.  By similar reasoning switching the roles of the
players, $o_{21}$ cannot be played with positive probability in
equilibrium.

But these two outcomes are also the two players' respective
most-preferred outcomes.  It is possible that player $1$ (resp.,
player $2$) assigns utility $1$ only to $o_{11}$ (resp., $o_{21}$) and
$0$ to all other outcomes -- in which case they would get utility $0$
on path, since their most-preferred outcomes are played with zero
probability on path.  It follows that if player $1$ (resp., player
$2$) is being punished, it must be by playing 100\% Right (resp.,
100\% Top) as otherwise player 1 (resp., 2) could get positive
expected utility in response.

However, then consider the case where player $1$ assigns utility $1$
only to $o_{11}$ and $o_{22}$, and $0$ to the others.  Because as just
discussed the punishment strategy must be 100\% Right, player $1$ is
able to get utility $1$ while being punished, by responding with
Bottom.  It follows that in equilibrium, she must also get utility
$1$, which means $o_{12}$ must be played with $0$ probability.
Reasoning similarly for player 2, $o_{22}$ must be also played with
$0$ probability.  We have now concluded that every outcome must
receive $0$ probability on path, which cannot happen.  So no robust
equilibrium exists.
\end{proof}

Note that only finitely many types are used in the proof, so this also
serves as an example of robust equilibrium nonexistence with finitely
many types.  Additionally, all the types used have all the utilities
set to $0$ or $1$.  As we will see, that is no accident; considering
only such extreme types is without loss of generality for the case of
total orders, and even partial orders (as long as these are orders on
pure outcomes).

It is quite common for games (without mediation or repetition) to have
pure Nash equilibria, sometimes even many (e.g.,~\cite{Rinott00:On}).
We next show that these correspond to robust equilibria as well.  To
determine whether something is a pure Nash equilibrium, one needs to
know only how the players ordinally compare the different outcomes of
the game, which is exactly what our formalism provides if we are
considering total orders.  The next definition formalizes this (and
also works for partial orders).

\begin{definition}
Define a {\em pure unmediated equilibrium} as a profile of actions for
the players, leading to outcome $o$, such that, if player $i$ can
unilaterally deviate to outcome $o'$, then it must be the case that $o
\succeq_i o'$.
\end{definition}

\begin{proposition}
Every outcome of a pure unmediated equilibrium can be sustained in a
robust equilibrium as well.
\label{prop:pure}
\end{proposition}
\begin{proof}
Simply always play the profile of the pure unmediated equilibrium,
whether punishing or not (i.e., no separate punishment strategy is
needed; one can set $q_{-i}=p_{-i}$).
\end{proof}

On the other hand, there are also games that have robust equilibria,
but all of them require mixed play, both on-path and in punishment.
The game in Figure~\ref{fi:MP2} is a simple example.
\begin{figure}
\begin{center}
\begin{tabular}{| c | c |}\hline
$o_{1}$ & $o_{2}$\\  \hline
$o_{2}$ & $o_{1}$\\  \hline
\end{tabular} \ \ \ \ \
\begin{tabular}{| c | c |}\hline
2, 1 & 1, 2\\  \hline
1, 2 & 2, 1\\  \hline
\end{tabular}
\end{center}
\caption{Left: together with the orders $o_{1} \succeq_1 o_{2}$ and
  $o_{2} \succeq_2 o_{1}$, this is an ordinal game.  Right: example
  utilities satisfying these constraints.}
\label{fi:MP2}
\end{figure}
Again this is a matching pennies game, but in this case players have
the same utilities for their two ``winning'' outcomes, as well as for
their ``losing'' outcomes, as in standard matching pennies.  Here, no
pure outcome can be sustained in robust equilibrium, but there is a
robust equilibrium where we randomize 50-50 over $o_1$ and $o_2$, and
each player's punishment strategy is to play 50-50 between the two
actions.

In other examples, robust equilibria require punishment play to be
different from on-path play.  A simple ordinal Prisoner's Dilemma
provides a quick example where the cooperative outcome can be
sustained only by defecting in the punishment phase -- though that
game additionally has a robust equilibrium that involves always
defecting.  (The ordinal game for Figure~\ref{fi:intro} is similar.)
The game in Figure~\ref{fi:PD2} is a richer version of the Prisoner's
Dilemma in which {\em any} robust equilibrium must use different
on-path and punishment strategies, and moreover these both must
randomize.
\begin{figure}
\begin{center}
\begin{tabular}{| c | c | c | c |}\hline
$o_{cc1}$ &  $o_{cc2}$ &  $o_{cd}$ &  $o_{cd}$\\  \hline
$o_{cc2}$ &  $o_{cc1}$ &  $o_{cd}$ &  $o_{cd}$\\  \hline
$o_{dc}$  &  $o_{dc}$  &  $o_{dd11}$ &  $o_{dd12}$\\  \hline
$o_{dc}$  &  $o_{dc}$  &  $o_{dd21}$ &  $o_{dd22}$\\  \hline
\end{tabular} \ \ 
\begin{tabular}{| c | c | c | c |}\hline
7, 4 & 4, 7 & 1, 8 & 1, 8\\  \hline
4, 7 & 7, 4 & 1, 8 & 1, 8\\  \hline
8, 1 & 8, 1 & 6, 2 & 3, 5\\  \hline
8, 1 & 8, 1 & 2, 6 & 5, 3\\  \hline
\end{tabular}
\end{center}
\caption{Left: together with the orders $o_{dc} \succeq_1 o_{cc1}
  \succeq_1 o_{dd11} \succeq_1 o_{dd22} \succeq_1 o_{cc2} \succeq_1
  o_{dd12} \succeq_1 o_{dd21} \succeq_1 o_{cd}$ and $o_{cd} \succeq_2
  o_{cc2} \succeq_2 o_{dd21} \succeq_2 o_{dd12} \succeq_2 o_{cc1}
  \succeq_2 o_{dd22} \succeq_2 o_{dd11} \succeq_2 o_{dc}$, this is an
  ordinal game.  Right: example utilities satisfying these
  constraints.}
\label{fi:PD2}
\end{figure}
This game has a robust equilibrium: on the path of play, randomize
50-50 between $o_{cc1}$ and $o_{cc2}$; and both players' punishment
strategies are to randomize 50-50 between the third and fourth
row/column.  This is an equilibrium because for player 1, if she
deviates that can only result in a distribution where 50\% of the
probability is on $\{o_{cd},o_{dd11},o_{dd22}\}$ and the other 50\% of
the probability is on $\{o_{cd},o_{dd12},o_{dd21}\}$.  Since she
weakly prefers $o_{cc1}$ to the former three, and $o_{cc2}$ to the
latter three, the deviation cannot make her better off.  A similar
argument shows the same for player $2$.

\begin{proposition}
For the game in Figure~\ref{fi:PD2}, in any robust equilibrium,
on-path strategies and punishment strategies must be different from
each other, and both must randomize.
\end{proposition}
\begin{proof}
(This proof uses similar reasoning as the proof of
  Proposition~\ref{prop:nonexistence}, so it is helpful to read that
  proof first.)  First, in robust equilibrium, on path, $o_{dc}$ and
  $o_{cd}$ can never be played, because they are each the worst
  outcome for one of the players, and that player can avoid that
  outcome by playing appropriately.  Second, this means that neither
  player ever receives its most-preferred outcome in equilibrium
  (on-path) play; and hence, neither the first two rows nor the first
  two columns can ever be played with positive probability in
  punishment strategies, because such a punishment strategy would
  allow the other player some chance of receiving their most-preferred
  outcome.  Third, there can be no robust equilibrium in which only
  the $dd$ outcomes receive positive probability on path.  This is
  because if we restrict the game to the strategies producing these
  outcomes, we obtain the game from Figure~\ref{fi:MP1} for which we
  have shown that no robust equilibria exist.  So, if a robust
  equilibrium in which only the $dd$ outcomes receive positive
  probability in on-path play did exist, it must be because either one
  of the first two rows or one of the first two columns receives
  positive probability in a punishment strategy, which we just showed
  is impossible.  Hence, the $cc$ outcomes must receive positive
  probability on path.  (In particular, this means that on-path play
  must be different from punishment play.)  Fourth, it cannot be the
  case that the only outcomes that receive positive probability in
  on-path play are the top four outcomes for one player (say, $o_{dc},
  o_{cc1}, o_{dd11}, o_{dd22}$, the top four for player $1$) because
  then the other player by playing uniformly at random always has some
  chance of getting an outcome preferred to all of those.  (In
  particular, this means that there must be some mixing in on-path
  play.)  Fifth, a pure punishment strategy (say, the third or fourth
  column) will not suffice because then the punished player can
  guarantee itself an outcome in its top four outcomes.
\end{proof}

The examples above can be verified using the linear program from the
section on Total Orders below.

\section{Finite Sets of Types}

In this section, we consider the setting where each player has only
finitely many types.  We show that this case can be solved with a
linear program that has a number of constraints that is linear in the
number of types.  While ordinal games generally have infinitely many
types, in later sections, we will show that in some cases, we can
actually restrict attention to a polynomial number of types; in
others, this is not true, but we can find a violated constraint in
polynomial time if one exists (i.e., a polynomial-time separation
oracle); whereas other cases yet are in fact coNP-hard to solve.

The following linear feasibility program (with variables $p(a)$ and
$q_{-i}(a_{-i}))$) describes a game's robust equilibria.
\begin{footnotesize}
\begin{align*}
& (\forall i, \theta_i, a_i' ) \ \sum_{a \in A} u_i(\theta_i, a) p(a) \geq \sum_{a_{-i} \in A_{-i}} u_i(\theta_i, a_i', a_{-i}) q_{-i}(a_{-i}) \\
&\sum_{a \in A} p(a) = 1 ; \ \ \ (\forall i) \sum_{a_{-i} \in A_{-i}} q_{-i}(a_{-i}) = 1 \\
& (\forall a \in A) \ p(a) \geq 0 ; \ \ \ (\forall i, a_{-i} \in A_{-i}) \  q_{-i}(a_{-i}) \geq 0
\end{align*}
\end{footnotesize}
Whether this linear feasibility program has a solution corresponds to
whether the answer to {\sc EORE} is positive.  We may add an objective
to obtain a linear program; doing so solves {\sc OMIRE}, but also {\sc
  SIRE}, as we may find an equilibrium that maximizes the probability
placed on any one particular profile $a^* \in A$ by simply adding an
objective of: {\footnotesize $\text{maximize } p(a^*)$.}  Finally, we
can also fix $p$ to some specific $p^*$ to solve {\sc AARE}.  (We will
use ``Linear Program 1'' to refer to all these linear programs
collectively.)  Thus:

\begin{theorem}
For games with finitely many types that are explicitly enumerated in
the input, {\sc EORE}, {\sc SIRE}, {\sc AARE}, and {\sc OMIRE} can be
solved using a single polynomial-sized LP, with
$O(|A|+\sum_{i}|A_{-i}|)$ variables and $O(\sum_{i} |\Theta_i| \cdot
|A_i|)$ constraints.
\end{theorem}

\section{Total Orders}

We now show how to efficiently solve for robust equilibrium in ordinal
games with total orders.

\begin{definition}
Given a total order $\succeq_i$ over outcomes, we say that
distribution $p$ over outcomes {\em stochastically dominates}
distribution $q$ over outcomes if for every outcome $o$, $\sum_{o': o'
  \succeq_i o} p(o') \geq \sum_{o': o' \succeq_i o} q(o')$.
\end{definition}

\begin{lemma}
$p$ and $(q_{-i})_{i}$ constitute a robust equilibrium if and only if
  for every $i$ and every $a_i'$, $p$ stochastically dominates
  $(q_{-i},a_i')$ for $\succeq_i$.
\label{le:stochdom}
\end{lemma}
\begin{proof}
For the ``if'' direction, if $p$ stochastically dominates
$(q_{-i},a_i')$, then it is possible to obtain $p$ from
$(q_{-i},a_i')$ by shifting probability towards outcomes that are more
preferred by $i$; therefore for every $\theta_i$, the expected utility
$i$ receives from $p$ must be at least as high as from
$(q_{-i},a_i')$. So we have a robust equilibrium.

For the ``only if'' direction, suppose there exists some $i$, $a_i'$,
and $o$ such that $\sum_{o': o' \succeq_i o} p(o') < \sum_{o': o'
  \succeq_i o} (q_{-i},a_i')(o')$.  Consider the type $\theta_i$ where
$u_i(\theta_i,o')=1$ if $o' \succeq_i o$ and $u_i(\theta_i,o')=0$
otherwise.  Then, $i$'s expected utility for not deviating is
$\sum_{o': o' \succeq_i o} p(o')$, whereas deviating to $a_i'$ gives
expected utility $\sum_{o': o' \succeq_i o} (q_{-i},a_i')(o')$; by
assumption the latter is larger, so we do not have a robust
equilibrium.
\end{proof}

Hence, the first (incentive) constraint in Linear Program 1 can be replaced by
$$(\forall i, o \in O, a_i' \in A_i) \mathop{\sum_{a \in A:}}_{o(a)
  \succeq_i o} p(a) \geq \mathop{\sum_{a_{-i} \in
    A_{-i}:}}_{o(a_i',a_{-i}) \succeq_i o} q_{-i}(a_{-i})$$ This
constraint can be interpreted as a restriction of the original
constraint -- specifically, restricting our attention to types for
which all utilities are either $0$ or $1$.  (Given that we are
considering total orders, those are types $\theta_i$ for which for
some $o$, $u_i(\theta_i,o')=1$ if $o' \succeq_i o$ and
$u_i(\theta_i,o')=0$ otherwise.)  A generalization of this result that
also works for partial orders is given as Lemma~\ref{le:zeroone}, with
a direct proof.

\begin{theorem}
For ordinal games with total orders, {\sc EORE}, {\sc SIRE}, {\sc
  AARE}, and {\sc OMIRE} can be solved using a single polynomial-sized
LP, with $O(|A|+\sum_{i}|A_{-i}|)$ variables and $O(|O| \cdot \sum_{i}
|A_i|)$ constraints.
\end{theorem}

\section{Partial Orders}

We now consider the case where players have only partial orders (over
pure outcomes), resulting in larger spaces of types consistent with
those orders.  We begin with the promised generalization of
Lemma~\ref{le:stochdom}.

\begin{lemma}
Given $p, (q_{-i})_i$ and a partial order $\succeq_i$ over outcomes,
if there is a type $\theta_i$ for which $i$'s incentive constraint is
violated, then there is also such a type $\theta_i'$ under which $i$
receives utility either $0$ or $1$ for every outcome.
\label{le:zeroone}
\end{lemma}
\begin{proof}
Consider some type $\theta_i$ for which $i$'s incentive constraint is
violated for some $a_i'$; let $t(\theta_i)$ denote the number of
distinct utility values strictly between $0$ and $1$ under $\theta_i$.
For example, if the utilities that $i$ receives for the respective
outcomes in the game under $\theta_i$ are $0, 0, 0, 1/4, 1/4, 1/2, 1,
1$, then $t(\theta_i)=2$ (because $1/4$ and $1/2$ are the two distinct
intermediate utilities).  For the sake of contradiction, suppose
$t(\theta_i)>0$ and moreover that this number is minimal among types
for which $i$'s incentive constraint is violated.  Then, take one of
these intermediate values -- call it $r$ -- and consider all the
outcomes that under $\theta_i$ have utility $r$.  If we increase or
decrease the utility for {\em all} these outcomes by the same amount
(say, to $r+\epsilon$ where $\epsilon$ is possibly negative), then all
the partial order constraints will still be satisfied, at least up
until the point where $r+\epsilon$ becomes equal to some other
intermediate or 0/1 value.  Moreover, {\em either} decreasing {\em or}
increasing them all by the same amount must keep the incentive
constraint violated, because the constraint is linear in the
utilities.  So, we can either increase or decrease them all up to the
point where they become equal to some other utility (whether
intermediate or in $\{0,1\}$), at which point we have found a type
$\theta_i'$ for which $i$'s incentive constraint for $a_i'$ is
violated with $t(\theta_i') = t(\theta_i)-1$.  But this contradicts
the supposed minimality of $t(\theta_i)$, proving the result.
\end{proof}

One challenge is that for a partial order, there can still be
exponentially many types with 0/1 utilities that satisfy the ordering
constraints.  Thus, unlike the case of total orders (where there are
only linearly many such types), we cannot write down one constraint
for each of these types.  Instead, we need to find an algorithm for,
given a candidate solution $p, (q_{-i})_i$ to Linear Program 1, {\em
  generating} a violated constraint if one exists -- i.e., a
separation oracle.  As it turns out, this can be done in polynomial
time.

\begin{lemma}
For partial orders over pure outcomes, the separation oracle problem
for Linear Program 1 can be solved in polynomial time, via $\sum_i
|A_i|$ max-flow problem instances that each have $O(|O|)$ vertices and
a number of edges that is on the order of the number of pairwise
comparisons in $\succeq_i$ (which is at most $|O|^2$).
\label{le:partial}
\end{lemma}
\begin{proof}
For finding a violated incentive constraint given values for $p$ and
$(q_{-i})_i$, we can iterate over all players $i$ and all $a_i' \in
A_i$.  The problem then is to find a $\theta_i \in \Theta_i$ such that
$$\sum_{a \in A} p(a) u_i(\theta_i, a) < \sum_{a_{-i} \in A_{-i}}
q_{-i}(a_{-i}) u_i(\theta_i, a_i', a_{-i})$$ We will do so by
maximizing the difference between the right- and left-hand sides,
i.e.,
$$\text{maximize}_{\theta_i \in \Theta_i} \sum_{a \in A}
(q_{-i}(a_{-i}) \boldone[a_i=a_i'] - p(a)) u_i(\theta_i, a)$$ where
$\boldone[a_i=a_i']$ returns 1 if $a_i=a_i'$ and $0$ otherwise.  Note
that in this separation oracle problem, the $p(a)$ and
$q_{-i}(a_{-i})$ are {\em parameters}, and the $u_i(\theta_i, a)$ are
{\em variables}.  In fact, since two action profiles that lead to the
same outcome $o$ must give the same utility, we can use variables
$u_i(\theta_i, o)$.  These variables have to satisfy the ordering
constraints, i.e., if we have $o \succeq_i o'$ then we must set
$u_i(\theta_i, o) \geq u_i(\theta_i, o')$.  Equivalently, if we (WLOG,
given Lemma~\ref{le:zeroone}) set all these variables to $0$ or $1$,
we must have $u_i(\theta_i, o') = 1 \Rightarrow u_i(\theta_i, o) = 1$.
This problem can be reduced to a problem in automated mechanism design
with partial verification~\cite{Zhang21:Automated} that they show can
be solved via a max-flow problem.  Specifically, in that problem,
there is a set of examples that must be classified as accepted or
rejected, but there are pairwise constraints of the form $\eta
\rightarrow \eta'$, meaning that if $\eta$ is accepted then $\eta'$
must be too.  There is a valuation function $v$ that assigns a value
$v(\eta)$ to each example $\eta$ (possibly negative), and the goal is
to maximize the sum of the values of the accepted examples.  We can
reduce to this problem by creating an example $\eta_o$ for each $o \in
O$, adding a constraint $\eta_{o'} \rightarrow \eta_{o}$ whenever $o
\succeq_i o'$, and setting $v(\eta_o) = \sum_{a \in A: o(a)=o}
(q_{-i}(a_{-i}) \boldone[a_i=a_i'] - p(a))$.
\end{proof}

\begin{theorem}
For ordinal games with partial orders (over pure outcomes), {\sc
  EORE}, {\sc SIRE}, {\sc AARE}, and {\sc OMIRE} can be solved using a
single LP whose separation oracle problem can be solved via $\sum_i
|A_i|$ max-flow problem instances that each have $O(|O|)$ vertices and
a number of edges that is on the order of the number of pairwise
comparisons in $\succeq_i$ (which is at most $|O|^2$).
\end{theorem}

\section{Partial Orders over Outcome Distributions}

More generally still, we may have not only ordering constraints over
pure outcomes, but also over distributions over outcomes.  For
example, we may have a previous observation where a player (weakly)
preferred one distribution over outcomes to another, and at least in
some cases we may be confident that this preference will persist to
future decisions.

In this section, we show that even for this case, we can obtain a
polynomial-time separation oracle, though it will no longer suffice to
restrict attention to utilities in $\{0,1\}$, as we will show next,
and we need a general linear program for the separation oracle rather
than just a max-flow problem.

Consider the game in Figure~\ref{fi:mixed}.
\begin{figure}
\begin{center}
\begin{tabular}{| c | c | c |}\hline
$o_{11}$ &  $o_{12}$ &  $o_{13}$\\  \hline
$o_{21}$ &  $o_{22}$ &  $o_{23}$\\  \hline
\end{tabular} \ \ \ \ \
\begin{tabular}{| c | c | c | c |}\hline
0.45, 0 & 0, 0 & 0, 0\\  \hline
0, 0 & 0, 0 & 1, 0 \\  \hline
\end{tabular}
\end{center}
\caption{Left: together with the ordering constraints $o_{11}
  \succeq_1 o_{22}$ and $o_{11} \succeq_1 (0.6o_{22} + 0.4o_{23})$,
  this is an ordinal game.  Right: example utilities satisfying these
  constraints.}
\label{fi:mixed}
\end{figure}
Suppose we consider $p_{\text{Top,Left}}=1$ and $q_{-1}(\text{Center})
= q_{-1}(\text{Right}) = 1/2$, and we are evaluating whether there are
any utilities (i.e., some type $\theta_1$ satisfying the ordering
constraints) for which player 1 would deviate and play Down instead.
0/1 utilities will not provide such a solution: either
$u_1(\theta_1,o_{11})=0$ but then we must also have
$u_1(\theta_1,o_{22})= u_1(\theta_1,o_{23})=0$, or
$u_1(\theta_1,o_{11})=1$; and in either case, player $1$ has no
incentive to deviate.  But the utilities given on the right of
Figure~\ref{fi:mixed} also satisfy the constraints, and under those
utilities player 1 has an incentive to deviate to Down given the above
$p$ and $q_{-1}$.

\begin{lemma}
  For partial orders over outcome distributions, the separation oracle
  problem for Linear Program 1 can be solved in polynomial time, via
  $\sum_i |A_i|$ linear programs that each have $O(|O|)$ variables and
  a number of constraints that is on the order of the number of
  comparisons in $\succeq_i$.
\end{lemma}
\begin{proof}
As in the proof of Lemma~\ref{le:partial}, the objective of the separation oracle problem can be written as
$$\text{maximize}_{\theta_i \in \Theta_i} \sum_{a \in A}
(q_{-i}(a_{-i}) \boldone[a_i=a_i'] - p(a)) u_i(\theta_i, a)$$ Unlike
in that proof, given the example above, here we cannot restrict the
$u_i(\theta_i, o)$ to take values in $\{0,1\}$ and instead need to let
them take values in $[0,1]$.  They also need to satisfy the partial
order constraints, where if we know that, for two probability
distributions over outcomes $r_1, r_2: O \rightarrow [0,1]$ it holds
that $r_1 \succeq_i r_2$, then we must have that $\sum_{o \in O}
r_1(o) u_i(\theta_i, o) \geq \sum_{o \in O} r_2(o) u_i(\theta_i, o)$.
But these are all linear constraints and a linear objective (because
the pairs $(r_1, r_2)$ are part of the input, and, in this separation
oracle problem, so are $p$ and the $(q_{-i})_i$).
\end{proof}
  
\begin{theorem}
For ordinal games with partial orders over outcome distributions, {\sc
  EORE}, {\sc SIRE}, {\sc AARE}, and {\sc OMIRE} can be solved using a
single LP whose separation oracle problem can be solved via $\sum_i
|A_i|$ linear programs that each have $O(|O|)$ variables and a number
of constraints that is on the order of the number of comparisons in
$\succeq_i$.
\end{theorem}

\section{Richer Constraints on Utilities}

In this section, we will see that some types of ordering constraints
do in fact lead to computational hardness, but we need a more
expressive language for that.  Specifically, we consider pairwise
orderings with logical connectives, in conjunctive normal form.  An
example preference-CNF formula is $((o_1 \succeq o_2) \lor (o_1
\succeq o_3)) \land ((o_3 \succeq o_4) \lor (o_4 \succeq o_1))$; the
guarantee we have on the agent's utilities is that the formula will
evaluate to {\em true}.  For example, the above formula does not allow
a type $\theta_i$ with utilities $u(\theta_i,o_1)=0,
u(\theta_i,o_2)=u(\theta_i,o_3)=1$ because the first clause would not
be satisfied.  However, every such formula is satisfiable by setting
all utilities to the same value.

\begin{theorem}
  With two players where player $1$'s type space is defined by a
  preference-CNF formula, {\sc EORE}, {\sc SIRE}, {\sc AARE}, and {\sc
    OMIRE} are each coNP-hard.  This is true regardless of what player
  $2$'s type space is.
\end{theorem}

\begin{proof}
Given an instance of SAT with variables $x_1, \ldots, x_m$, convert it
to a preference-CNF formula by creating, for every variable, an
outcome $o(x_i)$; as well as two additional outcomes $o_0$ and $o_1$.
Then, from the CNF formula in the SAT instance, replace each literal
$+x_i$ with $o(x_i) \succeq o_1$; and each literal $-x_i$ with $o_0
\succeq o(x_i)$.  For example, $(+x_1 \lor -x_2) \land (-x_1)$ would
result in the preference-CNF formula $((o(x_1) \succeq o_1) \lor (o_0
\succeq o(x_2))) \land (o_0 \succeq o(x_1))$.  One way to satisfy such
a preference-CNF formula is to set the utility for $o_0$ at least as
high as the utility for every $o(x_i)$, and the utility for $o_1$ at
most as high as the utility for every $o(x_i)$.  However, if we set
the utility for $o_0$ to be {\em strictly lower} than that for $o_1$,
then for each $x_i$, we must choose whether to set the utility for
$o(x_i)$ at least as high as that for $o_1$ (corresponding to setting
$x_i$ to {\em true}) or at most as high as that for $o_0$
(corresponding to setting $x_i$ to {\em false}); both cannot
simultaneously be true. Thus, under the condition that the utility for
$o_0$ is strictly lower than that for $o_1$, a clause in the
preference-CNF formula is true for a utility assignment if and only if
in the corresponding assignment for the original SAT instance, at
least one of the literals in the clause is set to {\em true}.  It
follows that the preference-CNF formula can be satisfied in a way such
that the utility for $o_0$ is strictly lower than that for $o_1$, if
and only if the original SAT instance is satisfiable.

Now consider the game in Figure~\ref{fi:hard}.
\begin{figure}
\begin{center}
\begin{tabular}{| c | c |}\hline
$o_{0}$ &  $o_{0}$  \\  \hline
$o_{1}$ &  $o_{1}$  \\  \hline
$o_{1}$  &  $o(x_1)$  \\  \hline
  $o_{1}$  &  $o(x_2)$ \\  \hline
  $\vdots$ & $\vdots$\\  \hline
    $o_{1}$  &  $o(x_m)$\\  \hline  
\end{tabular} \ \ \ \ \
\begin{tabular}{| c | c | c | c |}\hline
0, 0 & 0, 0\\  \hline
1, 0 & 1, 0\\  \hline
1, 0 & 0, 0\\  \hline
1, 0 & 0, 0\\  \hline
$\vdots$ & $\vdots$\\  \hline
1, 0 & 0, 0\\  \hline
\end{tabular} \ \ \ \ \
\begin{tabular}{| c | c | c | c |}\hline
1, 0 & 1, 0\\  \hline
0, 0 & 0, 0\\  \hline
0, 0 & 0, 0\\  \hline
0, 0 & 0, 0\\  \hline
$\vdots$ & $\vdots$\\  \hline
0,0 & 0,0\\  \hline
\end{tabular} 
\end{center}
\caption{Left: together with the preference-CNF formula that replaces
  $+x_i$ with $o(x_i) \succeq_1 o_1$ and $-x_i$ with $o_0 \succeq_i
  o(x_i)$ in the original SAT instance, this is an ordinal game.
  Center: example utilities satisfying these constraints if the SAT
  instance is satisfiable by setting all variables to {\em false}.
  Right: example utilities satisfying these constraints regardless of
  the SAT instance.}
\label{fi:hard}
\end{figure}
Suppose that player 1's utilities must satisfy the preference-CNF
formula above.  Then putting all the probability of $p$ on the top
left outcome can be sustained in robust equilibrium (and indeed the
game has a robust equilibrium at all) if and only if the original SAT
formula is {\em not} satisfiable.\footnote{So, SIRE's hardness follows
  from setting $a^*$ in SIRE to that top-left outcome (if not
  satisfiable, we can even put {\em all} the probability on $a^*$, and
  if satisfiable, there exists no robust equilibrium with $p(a^*)>0$
  because no robust equilibrium exists at all).  Similarly, for AARE
  we can just set $p^*(a^*)=1$.} This is because, if it is not
satisfiable, then there is no type $\theta_1$ such that
$u_1(\theta_1,o_1) > u_1(\theta_1,o_0)$, and this makes the top left
entry a pure Nash equilibrium, which by Proposition~\ref{prop:pure} is
sustainable in robust equilibrium.  On the other hand, if the original
SAT formula is satisfiable, then there is no robust equilibrium at
all, for the following reasons.  First, because it is satisfiable,
there exists a type $\theta_1$ for player 1 such that $1 =
u_1(\theta_1,o_1) > u_1(\theta_1,o_0) = 0$ (see center of
Figure~\ref{fi:hard} for an illustration).  Because player 1 can
guarantee herself an outcome of $o_1$, it follows that $o_0$ can get
no probability on the path of play in a robust equilibrium.  On the
other hand, whether the formula is satisfiable or not, there exists a
type $\theta_1$ for player 1 such that $1 = u_1(\theta_1,o_0) >
u_1(\theta_1,o_1) = u_1(\theta_1,o(x_1)) = \ldots =
u_1(\theta_1,o(x_m)) = 0$ (see right of Figure~\ref{fi:hard} for an
illustration).  Because player 1 can guarantee herself an outcome of
$o_0$, it follows that none of the other outcomes can get any
probability on the path of play, either.  But this means no robust
equilibrium exists.
\end{proof}

\section{When Are the Robust Equilibria Considered Here without Loss of Generality?}

The equilibria considered in this paper are robust not only to
uncertainty over utilities, but also to modeling assumptions, in the
sense that they will remain equilibria under a variety of models as
discussed in the introduction.  Nevertheless, we may wonder whether
under certain assumptions, additional equilibria (that remain robust
to uncertainty over utilities) become available.  This is what we
investigate in this section, showing that in some cases no other
equilibria are possible, but in other cases, if we are confident some
properties hold, there may be additional equilibria of a different
type.

The type of equilibrium studied in this paper precisely fits the
mediated-equilibrium model.  This is because the mediator must specify
a distribution $p$ over outcomes (corresponding to a correlated
strategy for all players), as well as for each $i$, a correlated
strategy $q_{-i}$ for using the other players' actions to punish
player $i$ if that player deviates.

For the case of repeated games, one may wonder whether we can use the
fact that we play multiple rounds in the punishment stage, by
punishing some types in some rounds and other types in other rounds
(as opposed to using the same $q_{-i}$ in every round).  If we were
only allowed to play pure actions in the punishment phase, this is in
fact correct: if action 1 were effective for punishing type 1, and
action 2 for punishing type 2, then we may wish to play action 1 in
odd rounds and action 2 in even rounds, so that neither type would get
high utility and consequently neither type has an incentive to
deviate.  On the other hand, if we are allowed mixed actions, then of
course we may just as well play each action with probability $1/2$
each round.  Indeed, generally:

\begin{proposition}
Consider a planned sequence of punishment mixed actions $q_{-i}^1,
q_{-i}^2, \ldots, q_{-i}^m$ (where the superscript indicates the round
in which the punishment is to be played).  Let $q_{-i}^* = (q_{-i}^1 +
q_{-i}^2 + \ldots + q_{-i}^m) / m$ be the average of these mixed
actions.  Then against any type $\theta_i$, repeating $q_{-i}^*$ for
$m$ rounds is at least as effective as the original sequence (i.e.,
$\theta_i$ cannot achieve higher utility against the new sequence).
\end{proposition}
\begin{proof}
  Consider an alternative scenario where the original sequence is
  used, but player $i$ has imperfect recall and cannot remember which
  round it is.  Then player $i$ is naturally modeled as having a
  uniform belief over the index of the current round, so its belief
  about the joint action of players $-i$ in the current round is
  captured by $(q_{-i}^1 + q_{-i}^2 + \ldots + q_{-i}^m) / m =
  q_{-i}^*$, and so $i$ should best-respond to that.  The loss of
  information from imperfect recall cannot have made $i$ better off,
  so having to best-respond to $q_{-i}^*$ a total of $m$ times cannot
  be better for $i$ than having to best-respond to the original
  sequence, no matter its type.
\end{proof}

However, the above result only holds if the sequence of punishment
mixed actions is preplanned.  Can we do better by dynamically adapting
to the punished player's actions, as these reveal information about
the punished player's type?  For this to make sense, first of all the
punished player's actions must be {\em observable}, and also the
punished player's type needs to {\em persist} (not change) from round
to round.  If {\em both} these conditions hold in a repeated game,
then in fact the type of equilibrium studied in this paper is {\em
  not} without loss of generality, i.e., punishment strategies that
condition on past behavior by the punished player can in fact be more
effective than punishment strategies that do not do so.  The following
example illustrates this.

\begin{example}
Consider player 1 (whom player 2 is trying to punish), who has one of
three types: $r$ (red), $b$ (blue), or $g$ (green).  She also has
three actions, $r$, $b$, or $g$; and player 2 has the same three
actions.  Player 1 receives a utility of $1$ whenever she plays the
color corresponding to her type {\em and} player 2 plays a different
color, and $0$ otherwise.  For any fixed mixed action $q_2$ of player
2, there will be at least one color to which $q_2$ assigns probability
at most $1/3$.  Thus, if player 1 has the type corresponding to that
color, then player 1 can achieve an expected utility of at least $2/3$
against that mixed action.  However, if this is a repeated game,
player 2 can observe player 1's actions, {\em and} player 1's type
persists over time, then player 2 can punish player 1 better by using
the following strategy: play whichever color player 1 played last
round.  This guarantees that every round in which player 1 obtained
utility $1$ (which must have been by playing the color of her type) is
followed by one in which she gets utility $0$ (as player 2 will be
playing the color of her type), so she can achieve a utility of at
most $1/2$ on average per round.
\end{example}

Indeed, significant attention has already been paid to
infinitely-repeated two-player zero-sum games in which one player has
private information about a persistent state.  These games are known
to be equivalent to a particular zero-sum public-signaling game, and
public signaling games are computationally
intractable~\cite{Aumann95:Repeated,Sorin02:First,Dughmi19:On,Conitzer20:Bayesian}.
Studying those techniques in ordinal games is an interesting direction
for future work.

\section{Other Future Research}

We are still left to deal with the fact that robust equilibria do not
always exist, even with total orders.  The approach in this paper
remains valuable as in many cases such equilibria do exist, and in
those cases they seem to be highly desirable from the perspectives of
robustness and safety.  Indeed, it seems that in practice, often,
while we do not know exactly by how much a player prefers one outcome
to another, we know enough to construct punishment strategies that
will discourage undesired behavior.  Still, it would be desirable to
identify precise sufficient conditions for robust equilibria to exist.
One natural candidate would be games that require some coordination
among the players.  For example, consider a game in which agents must
choose a location in which to go work.  They benefit from being in the
same location with other cooperative agents, but would suffer from the
presence of defecting agents.  Then, it is natural to punish a
defecting agent by having the other agents coordinate on a randomly
chosen location, without telling the punished agent that location
(cf.~the use of social exclusion as punishment~\cite{Hadfield13:Law}).

Additionally, it would be desirable to develop techniques for dealing
with games in which no robust equilibria exist.  Such techniques will
necessarily sacrifice some of the properties of robust equilibria as
described here, but they may still be well-motivated.  As we discussed
in the previous section, in repeated games with observable actions and
persistent types, additional solutions are sometimes possible.
Another approach is to allow agents to {\em report} their types to the
mediator (or otherwise communicate about their types), thereby moving
us towards a mechanism design setting.  (See
also~\cite{Forges13:Folk,DiGiovanni23:Commitment}.)

More broadly, the study of robust solutions in game theory is
important for keeping systems of multiple self-interested agents safe
and cooperative.  As we have argued in this paper, settings that
enable folk theorems -- which may be especially common in the context
of AI agents, e.g., program games~\cite{Tennenholtz04:Program} -- are
better suited to such robust solutions than traditional settings where
we, say, compute a Nash equilibrium of a normal-form game.

\newpage

\section*{Acknowledgments}

I thank the Cooperative AI Foundation, Polaris Ventures (formerly the
Center for Emerging Risk Research), and Jaan Tallinn's donor-advised
fund at Founders Pledge for financial support.

\bibliography{/afs/cs.cmu.edu/user/conitzer/a/references/references.bib}

\end{document}